\documentstyle[twocolumn,aps]{revtex}

\begin{document}
\twocolumn[\hsize\textwidth\columnwidth\hsize\csname
@twocolumnfalse\endcsname

\title{ Spin, charge and orbital ordering in La$_{0.5}$Sr$_{1.5}$MnO$_4$}
\author{Priya Mahadevan$^1$\cite{nrel},
K. Terakura$^2$ and D. D. Sarma$^3$\cite{jnc}}
\address{(1). JRCAT-ATP, Tsukuba, Ibaraki 305-0046, Japan \\ (2) JRCAT-NAIR,
Tsukuba, Ibaraki 305-8562, Japan \\
(3). SSCU, Indian Institute of Science, Bangalore-560012}
\date{\today}

\maketitle

\begin{abstract}
We have analyzed the experimental evidence of charge and orbital ordering
in La$_{0.5}$Sr$_{1.5}$MnO$_4$ using first
principles band structure calculations. Our results
suggest the presence of two types of Mn sites in the system.
One of the Mn sites behaves like an Mn$^{3+}$
ion, favoring a Jahn-Teller distortion of the surrounding oxygen atoms,
while the distortion around the other is not a simple breathing mode kind.
Band structure effects are found to dominate the experimental spectrum for
orbital {\it and charge ordering}, providing an alternate explanation
for the experimentally observed results.
\end{abstract}

\pacs{PACS number: 61.10.Dp, 71.20.-b}
]


In recent times there has been a resurgence of interest in the manganites
as a result of the wide range of physical properties that they exhibit. In
some of the doped compounds, there is a real space ordering
of the charge carriers in
certain orbitals, resulting in orbital ordering (OO) and 
sometimes charge ordering (CO).
As the magnetic and transport properties are closely correlated
with the orbital and charge degrees of freedom, there have been intense efforts -
both experimental and theoretical - to clarify
the microscopic mechanisms giving rise to such exotic phenomena\cite{co}.
While techniques such as electron microscopy and neutron scattering
have been traditionally used to detect CO as well as magnetic
ordering\cite{expt}, the direct observation of
OO has been difficult.
Murakami {\it et al.} performed pioneering experiments on
LaMnO$_3$ \cite{lamno3}
and La$_{0.5}$Sr$_{1.5}$MnO$_4$ \cite{lasrmn} and were successfully able
to detect OO.
The technique they used exploited the anisotropy of the x-ray
scattering tensor at an absorption edge, thereby making the intensity of the
scattered  x-rays sensitive to the valence electron distribution.

La$_{0.5}$Sr$_{1.5}$MnO$_4$ is a layered manganite with an average valence
of 3.5
at each Mn site. 
In the high temperature phase, above the CO temperature, all Mn-O
bond lengths within the {\it ab} plane are equal \cite{Immm}.
Neutron scattering studies by
Sternlieb {\it et al.} \cite{sternlieb}
indicated 
a real space ordering of two distinct Mn sites below 217~K, with the two 
sites identified
as Mn$^{3+}$ and Mn$^{4+}$ sites. The intensity variation observed at the
two sites
was interpreted as arising from a 1\% breathing mode-like movement of
the oxygen
atoms in the {\it ab} plane towards one set of Mn atoms, identified as the
higher valent Mn$^{4+}$ sites. 
Later experiments by Murakami {\it et al.} \cite{lasrmn}
used the anisotropy of the x-ray scattering tensor to show that the Mn sites
form a
pattern of alternating Mn$^{3+}$-Mn$^{4+}$ sites in the {\it ab} plane, 
with both CO and OO taking place at the same temperature, 217 K.
They claimed that the charge modulation between the two Mn sites is not small,
but corresponds to as much as an {\it integral} charge
fluctuation between neighboring Mn sites.
However, if this is the case,
strong correlations are necessary to localize charge carriers.
Surprisingly, however, certain other spectroscopic properties \cite{optical}
of this system
can be easily explained by a single particle approach.

The contrasting experimental evidence where one set of spectroscopic data
could be easily
explained by an effective single particle approach, while another set
implied the
importance of strong correlations in the system prompted us to perform a
detailed
analysis of the electronic structure using the
density-functional pseudopotential framework. Apart from providing a
microscopic mechanism for the observed {\it orbital and charge-ordering}
in this system, our results have important
implications on recent anomalous x-ray scattering
experiments on Pr$_{1-x}$Ca$_x$MnO$_3$~\cite{newexpt1}, LaSr$_2$Mn$_2$O$_7$~\cite{newexpt2}
and NaV$_2$O$_5$~\cite{newexpt3}.

The details of the computational method are described
elsewhere~\cite{sawada}, and we only mention minimum additional
aspects.  The cutoff energy for the plane wave expansion of
the eigen state is 20 Ry, being smaller than the value of
30.25 Ry in our usual calculations because of the large unit cell in
the present calculation.  The k-point mesh used is 4x4x2.  The
generalized gradient approximation (GGA) was used for the exchange
functional~\cite{perdew96}.  The
alloying effects of (La$_{0.5}$Sr$_{1.5}$) at the A-site of the lattice
are treated within the virtual
crystal approximation~\cite{lee98}.

The Mn spins order below 110~K
forming one-dimensional ferromagnetic zigzag chains in the
{\it ab} plane coupled antiferromagnetically to each other, known as the 
CE-type magnetic ordering. As the exchange
splitting is large (about 3 eV), the amplitude for electrons to hop from a zig-zag
chain of parallel spins (indicated by thick solid lines in Fig.~1a) to the neighboring
chain is small. Hence the electronic structure is governed by these one-dimensional
chains. Considering a single chain,
it was shown \cite{igor} that the anisotropic
hopping between the Mn $d$ orbitals could explain the OO at
the two Mn sites. We have performed {\it ab-initio} calculations considering the
complete 3-dimensional Immm structure \cite{Immm}.
Similar to the findings in ref. \cite{igor}, we also
found the kind of OO that is observed
experimentally. Choosing the x and y axes as shown in Fig.~1, we found that
while the $d_{3x^2-r^2}$ orbital was preferentially occupied on atom 1,
the $d_{3y^2-r^2}$ orbital was preferentially occupied on atom 3.
These results can be understood within the framework of a
simple nearest neighbour tight-binding model involving
the $e_g$ orbitals: $d_{3x^2-r^2}$ and $d_{y^2-z^2}$ on 
atom 1, $d_{3y^2-r^2}$ and $d_{z^2-x^2}$ on atom 3 as well as
$d_{3z^2-r^2}$ and $d_{x^2-y^2}$ on atoms 2 and 4 as used in Ref.~\cite{igor}.
While the $d_{3x^2-r^2}$ orbital at Mn(1) and $d_{3y^2-r^2}$ orbital on
Mn(3) hybridize with both $e_g$ orbitals on atoms 2 and 4, the $d_{y^2-z^2}$
on Mn(1) and $d_{z^2-x^2}$ on Mn(3) do not. The eigenvalue spectrum of such a simple 
model, consists of two bonding bands, energetically separated from four non-bonding bands, with
the antibonding bands at higher energies. As the average valence of Mn is 3.5, there are two electrons
in the $e_g$ orbitals of the 4 Mn atoms comprising the chain. 
Hence the two bonding bands which have dominantly
$d_{3x^2-r^2}$ character on atom 1 and $d_{3y^2-r^2}$ character on atom 3 are occupied. Thus, 
the one-dimensional chains which are a consequence of the magnetic structure drive the OO
within these calculations. However, the difficulty in approaching the problem this way is that 
the Immm structure exists only above the CO/OO temperature and there 
are extensive evidence in the literature suggesting a symmetry lowering below
the ordering transition. However, the details of the 
crystal structure of this compound is shrouded in controversy.
It was suggested~\cite{sternlieb} that the structure, 
within the charge 
ordered state,
is Cmmm with lattice constants equal to $\sqrt 2$ a, $\sqrt 2$ a, c where
a and c are the cell dimensions of the basic Immm cell. 
More recent work by Larochelle \cite{simon}, however,
indicate that the structure
is Ammm or one of its two subgroups Am2m or A222, with 
the lattice parameters being $\sqrt 2$a, 2$\sqrt 2$ a, 
and c. None of the structural 
investigations have been able to provide the atom positions, particularly
for the oxygen sites, owing to the large unit cell and the existing
data quality \cite{simon}. This work however suggests certain possible
distortions of the oxygens, though not in agreement with ref.~\cite{sternlieb}. 
In view of such uncertainities,
we have optimised the structure of this
system using {\it ab-initio} band structure calculations. The structure
that we find (see Fig. 1) 
is P2$_1$/m with lattice parameters
$\sqrt{2}$a, 2$\sqrt{2}$a and c. The lattice parameters as well as the space group 
that we have obtained are 
the same as in 
ref.~\cite{simon}, if the 
virtually negligible ($\sim$ 0.015 \AA) displacements of the Mn$^{4+}$
atoms shown in Fig~1 are ignored.

On optimizing the internal coordinates, the system
exhibited different distortions of the MnO$_6$ octahedra associated with
different Mn atoms. 
The direction of displacement of the oxygen
atoms in the ab-plane is indicated by the arrows shown in Fig.~1b, though 
the oxygen atoms have been left out of the figure for added clarity.
The Mn site, Mn(3),
showing $d_{3y^2-r^2}$ OO 
lowered its energy by an elongation
of the Mn-O bonds in the y-direction. This kind of distortion can be
understood easily within the framework of crystal-field effects. A tetragonal
distortion of a MnO$_6$ octahedron resulting in an elongation of the Mn-O
bondlengths in the  y-direction, lowers the bare energy of
the $d_{3y^2-r^2}$ orbital. Thus, the sites Mn(1) and Mn(3) behave like Mn$^{3+}$
species, sustaining a Jahn-Teller (JT) like
distortion of the surrounding oxygens
in two mutually perpendicular directions, giving rise to the orbital ordering. As a consequence,
the oxygen atoms surrounding the sites labelled Mn(2) and Mn(4) sustain 
distortions 
with the oxygen atoms along the x-axis being displaced in one 
direction  and the ones along the y-axis being displaced in 
another direction, as shown in the figure, 
in contrast to the suggestion in ref.~\cite{sternlieb} where
all four oxygen atoms surrounding the Mn$^{4+}$ site move 
closer to the Mn atom. 
We found the kind of distortion suggested in that 
work to have a higher energy. The sites Mn(2) and Mn(4)
have been identified as Mn$^{4+}$ sites in the literature and we retain
that nomenclature, although within our calculations the charge difference
between the so-called Mn$^{3+}$ and Mn$^{4+}$ species is negligible.

While our calculations suggest
different distortions of the surrounding oxygens about
the Mn$^{3+}$ and Mn$^{4+}$
sites, the actual magnitude of the distortion is fairly small~\cite{comment1}.
The underestimation of the magnitude of the JT distortion
by these first-principles approaches is well-known, though these
methods do get the nature of distortions
correctly. An example of this
is LaMnO$_3$ where the nature of distortion was
correctly predicted, while
the magnitude of the theoretical JT distortion was
found to be half of the
experimental value~\cite{sawada}.
We simulated the 
neutron diffraction pattern using the optimized 
coordinates as well as the coordinates assuming a Jahn-Teller distortion
equal to what is found in LaMnO$_3$. The simulated 
diffraction
patterns were nearly identical, 
indicating the difficulty of determining the oxygen positions 
with any precision from such measurements. Therefore
the enhanced distortion of the oxygen octahedra 
are not inconsistent with the structural 
data available so far. We in fact hope that the present theoretical 
considerations will help in the future structural investigations. It is also
significant that with optimized JT
distortion, the correct ground state magnetic order cannot be
reproduced for LaMnO$_3$~\cite{sawada}.
The situation seems to be the same in the present case,
where the CE type antiferromagnetic state is more stable 
than the ferromagnetic state only with the enhanced JT distortion,
consistent with earlier observations~\cite{mizokawa}. Hence in the
subsequent analysis, we have extrapolated the magnitude of the
JT distortion to the value observed in LaMnO$_3$.

The experiments of Murakami
{\it et al.} \cite{lasrmn} analyzed the energy dependence of the OO
superlattice reflection across the Mn K absorption edge.
The strong energy dependence was interpreted as arising from a splitting 
of $\sim$ 5~eV between the 4$p_x$ and 4$p_y$ PDOS at the sites Mn(1) and Mn(3) which
show OO. While the 4$p_x$ states were suggested to be 5~eV 
lower than the 4$p_y$ on one Mn atom, the order was reversed on the other
Mn atom.
Theoretically there are
two contrasting interpretations possible for such a splitting.
Ishihara and Maekawa~\cite{ishihara} argued in the context LaMnO$_3$ 
that the splitting in the Mn 4$p$
states induced by the intraatomic $p-d$ Coulomb interaction produces
such a strong tensor character of the scattering form factor.
However, it was pointed out~\cite{other_pap} that
as the 4p states are extended, the suggested $p$-$d$ Coulomb interaction
strength is unphysically large and the splitting in the 4$p$ states 
is actually caused by the JT
distortion, though Coulomb interactions, particularly within the Mn 3$d$ 
manifold, may have important consequences for other properties in such systems.
These suggestions are consistent with an earlier analysis of transition metal 
$K$-edge XAS in Fe and Co oxides~\cite{XAS}.

In Fig.~2 we plot the Mn 4$p_x$ and 4$p_y$ partial densities of states (DOS) in the
energy range 10-20 eV above the Fermi level, projected
onto the Mn site (Mn(1)) showing $d_{3x^2-r^2}$ ordering. A splitting
of $\sim$ 3~eV is clearly visible, with the 4$p_x$ states located at lower energies.
The order is reversed at the other Mn site (Mn(3)) which shows $d_{3y^2-r^2}$
ordering. Thus, our results on La$_{0.5}$Sr$_{1.5}$MnO$_4$ are consistent with 
the suggestion~\cite{other_pap} in the context of LaMnO$_3$ that the JT distortion
to be not only the driving force for the orbital ordering, but also responsible for 
the specific experimental effect of the pronounced tensor character of the form 
factor {\it via} the splitting of the 4$p$ partial DOS.

Strong enhancement in the
intensity at the energy corresponding to the absorption edge of Mn$^{3+}$
was observed at the CO superlattice 
reflection.~\cite{lasrmn}
These results suggested the presence of a second Mn atom in the system,
with its absorption edge 4~eV higher than that for the Mn$^{3+}$ atoms.
As the absorption edge for Mn$^{4+}$ atoms lies $\sim$4~eV above the
absorption edge for Mn$^{3+}$ atoms, they claimed that this was
a direct evidence of an ordering of the two
charge species - Mn$^{3+}$ and Mn$^{4+}$ in this system. 
From the present calculations, we find that as 
the environment for the Mn$^{3+}$-like
sites is different from the environment for the Mn$^{4+}$-like sites, there is
a substantial modification in the Mn 4$p$ PDOS, though the net charge
associated with these two sites, 4.53 at Mn 1 and 3, and 4.51 at Mn 2 and 4, 
are almost identical.  Hence, the modifications at the
so-called CO superlattice reflection cannot be
interpreted on the basis of a substantial difference in the charge state of
the two inequivalent sites. Interestingly, our results for the electronic structure
at these two sites provide a natural explanation for the observed results.
With the x- and y-axis defined in Fig.~1 we
have the 4$p_x$ states
at Mn$^{3+}$(1) at lower energies compared to the Mn 4$p_x$ states at Mn$^{4+}$
site (Fig.~3). As a result we have an intensity maximum at the CO
superlattice
reflection. If the axes are rotated by 45 degrees, one finds that the Mn
4$p_{x'}$
states at Mn$^{3+}$(1) and those at the Mn$^{4+}$ site almost
coincide (inset in Fig.~4). This results in a
minimum in the intensity. A similar angular dependence is also
observed at the OO superlattice reflection. Our
reinterpretation is supported by recent
experiments on Nd$_{0.5}$Sr$_{0.5}$MnO$_3$ \cite{ndsrmn} which also found
a similar angular dependence of the intensity at a CO
superlattice reflection.

In conclusion, we have carried out an analysis of the experimental 
observation for orbital and charge orderings in La$_{0.5}$Sr$_{1.5}$MnO$_4$.
Our results indicate the presence of two Mn species with very
different environments.
One of the Mn species has a JT distortion of the oxygens
surrounding that atom and hence, in this sense, behaves like an Mn$^{3+}$ species.
The distortions around the so-called Mn$^{4+}$ atoms are different from the simple
breathing mode
distortion suggested earlier.
Extrapolating the magnitude of the JT distortion around the Mn$^{3+}$
to equal the value observed in LaMnO$_3$ is necessary to obtain the
correct description of the ground magnetic state. Further,  the energy dependence at the
orbital ordering as well as the charge ordering superlattice
reflections can be explained
by the distortions of the MnO$_6$ octahedra.
The present results establish that
no real-space substantial charge ordering needs to be invoked in this
system in order to explain the experimentally
observed results; instead band structure effects
are responsible for the dependence of the superlattice 
intensities on the energy and the
orientation. The above interpretation is consistent 
with the observation that both charge ordering as
well as the orbital ordering show the same temperature dependence.
As the origin  of the intensity
variation observed at both reflections is the same, namely the lattice 
distortions, our results
explains why the charge ordering as well as the orbital ordering
transition occur over the same temperature range in this system.

The authors acknowledge Dr. Solovyev and Dr. Fang for valuable
discussions and comments.
The present work is partially supported by NEDO.

\begin{figure}
\caption{ (a). A schematic diagram of the spin, charge and orbital ordering of the Mn atoms
in the $ab$ plane. The magnetic structure consists of the zigzag chain (thick solid line
connecting atoms 1, 2, 3 and 4), coupled antiferromagnetically to the neighboring chains.
(b). The direction of displacements of the oxygen atoms when the atom positions are
relaxed are shown by the thick arrows, while the 
displacements of the Mn$^{4+}$ atoms are indicated by the small arrows. 
The two choices of coordinate axes x and y
and x$^{\prime}$ and y$^{\prime}$ used are shown.
}
\end{figure}

\begin{figure}
\caption{ The Mn 4$p_x$ (solid line) and 4$p_y$ (dotted line) partial DOS projected onto
the Mn$^{3+}$ site showing $d_{3x^2-r^2}$ ordering. 
}
\end{figure}

\begin{figure}
\caption{ The 4$p_x$ partial DOS at the Mn$^{3+}$ (solid line) and Mn$^{4+}$ sites (dotted line)
are shown for the x and y axes defined in Fig.~1, and in the 
inset  for x$^{\prime}$ and y$^{\prime}$ axes
defined in Fig.~1.
}
\end{figure}


\begin{references}
\bibitem{comment1} In the present structural optimization,
the longer Mn-O distance and the shorter one at sites 1 and 3
are 1.98 {\AA} and 1.90 {\AA}, respectively.  Similarly at sites 2
and 4, they are 1.89 {\AA} and 1.96 {\AA}.
\end{references}
\end{document}